\documentclass[review]{elsarticle}

\usepackage{lineno,hyperref}
\modulolinenumbers[5]
\biboptions{authoryear}

\usepackage{epsfig, amsmath, amsfonts, amssymb, amsthm,multirow,algorithm,algorithmic}
\usepackage{graphicx}
\usepackage{setspace}
\usepackage{enumerate}
\usepackage{lipsum}
\usepackage{epstopdf}
\usepackage{bm}
\usepackage{natbib}

\usepackage{hyperref}
\hypersetup{citecolor=blue,colorlinks=true}

\theoremstyle{plain}

\newcommand{\beq}{\begin{equation}}
\newcommand{\eeq}{\end{equation}}
\newcommand{\ber}{\begin{eqnarray}}
\newcommand{\eer}{\end{eqnarray}}

\newcommand{\bit}{\begin{itemize}}
\newcommand{\eit}{\end{itemize}}
\newcommand{\iid}{\stackrel{\mathrm{i.i.d.}}{\sim}}
\newcommand{\ben}{\begin{enumerate}}
\newcommand{\een}{\end{enumerate}}

\numberwithin{equation}{section}

\begin{document}

%
%

\begin{frontmatter}

\title{Spline-Based Bayesian Emulators for Large Scale Spatial Inverse Problems}

\author{Anirban Mondal}
\address{Case Western Reserve University, Cleveland, OH 44106, USA}

\author{Bani Mallick}
\address{Texas A\& M University, College Station, TX, 77843, USA}

\begin{abstract} 

A Bayesian approach to nonlinear inverse problems is considered where the unknown quantity (input) is a random spatial field. The forward model is complex and non-linear, therefore computationally expensive. An emulator-based methodology is developed, where the Bayesian multivariate adaptive regression splines (BMARS) are used to model the function that maps the inputs to the outputs. Discrete cosine transformation (DCT) is used for dimension reduction of the input spatial field. The posterior sampling is carried out using trans-dimensional Markov Chain Monte Carlo (MCMC) methods. Numerical results are presented by analyzing simulated as well as real data on hydrocarbon reservoir characterization.
\end{abstract}


\end{frontmatter}


\section{Introduction}
Mathematical models are often used in many areas of science and engineering to describe some physical process over a spatial field. Often these models are complex in nature and are described by a system of nonlinear ordinary or partial differential equations, which cannot be solved analytically. Given the input spatial field and some other model input parameters, such systems can be solved numerically by running a complex computer code, providing the corresponding outputs. These computer codes, usually called the ``forward model'' or ``forward simulator'', generally use a discretized approximation to the ordinary or partial differential equations. The spatial inverse problem consists of inferring about the unknown spatial field given the output observations. Here we focus on the Bayesian approach which casts the inverse solution as a posterior distribution of the unknown spatial field given the outputs. The Bayesian approach also contains a natural mechanism for regularization in the form of prior information. The posterior distribution of the inputs, in most of the cases, is intractable so Markov Chain Monte Carlo (MCMC) methods are often used to sample from the posterior. However, the likelihood term in the posterior distribution contains the ``forward simulator" which is governed by a system of nonlinear ordinary or partial differential equations. The corresponding numerical solver is often very expensive in terms of CPU time required even for a single run. Furthermore, the Monte Carlo sampling approach relies on a large number of execution of the likelihood, as thousands of samples from the posterior distribution are needed, which makes the Bayesian approach computationally very challenging.
 
Several attempts to accelerate the Bayesian inference in inverse problems have been made as alternatives to the direct Markov Chain Monte Carlo technique. For example, \cite{ehl05} and \cite{mondal2} have used a two-stage MCMC method, where the proposals are screened at the first stage using a much cheaper forward solver on a coarser grid. A substantial amount of CPU time is saved by rejecting the bad proposals in the first stage. However, the expensive forward simulator over the fine-grid still needs to be computed for a large number of iterations when the proposals are accepted in the first stage.

Another very popular approach is based on emulation of the simulation outputs which offers substantial efficiency gain over standard methods (see \cite{sacks}, \cite{ken}, \cite{ohag2}, \cite{hig}, \cite{oak}). An emulator is a statistical approximation of the forward model that is constructed by using training samples of simulation runs. Uncertainty and sensitivity analysis can be handled easily using an emulator as its execution is essentially instantaneous.
Given a set of training runs of the forward model for some suitably chosen inputs, also called the design points, both the inputs and the outputs are used to estimate the unknown complex function that maps inputs to the outputs.
The efficiency gains arise because it is usually possible to emulate the simulator output to a high degree of precision using only a few hundred runs of the simulator.

In the statistical emulator approach, the computer simulation model is treated as a ``black box", which is usually modeled by a Gaussian Process (GP). Generally, two groups of input model parameters are considered. One group comprises unknown context-specific inputs, which are calibrated using output observations from the physical process. These are called calibration inputs. The other group comprises all the other model inputs that are assumed to be known for the corresponding output observations.

In this article, we consider a spatial inverse problem where the unknown calibration input of the physical process is a high dimensional spatial field. Our goal is to calibrate an unknown spatial field, whereas most of the statistics literature in this area focuses on calibrating a set of unknown parameters. More specifically, we would like to infer about the input spatial process $Y(s,\omega), (\omega\in \Omega, s\in D \subset \mathcal{R}^2)$ in a very fine grid, given the observed output data and the other known model inputs. Here the spatial process is defined over the sample space $\Omega$ and the closed domain $D$. In addition, we also consider a number of independent simulation runs of the forward model, where the specified known model inputs, the input spatial field, and the corresponding simulator outputs are completely known. Our goal is twofold: (i) to approximate the forward simulator by an emulator based on these simulated and as well as real data; (ii) to infer about the unknown spatial field on a very fine grid given the output observations. Some observations on the spatial field on a coarser grid and very few available observations in the fine grid may also be included in the inferential procedure using a Bayesian hierarchical model.

While the majority of the literature uses GP-based emulators, here we use an alternative specification based on splines. In particular, we use an emulator based on the Bayesian approach to multivariate adaptive regression splines (BMARS), as introduced by \cite{den1}. Gaussian process emulators are usually suitable for interpolating smooth surfaces but may fail to handle more complex non-stationary surfaces. The multivariate adaptive regression splines approach has the ability to interpolate complex, non-stationary surfaces. A difficulty arises while implementing BMARS as in our situation the regressor is a high dimensional spatial field. Hence, we propose to use a truncated discrete cosine transformation of the spatial field where the original process $Y(s,\omega)$ is represented by a finite low dimensional parametrization. The finite-dimensional transformed DCT coefficients and other known model input parameters are then used as the regressors while fitting the BMARS for the unknown forward simulator. The inputs for the simulation data are generated using the Latin hypercube design. Furthermore, the validation of the computer code is done by fitting the BMARS model on a training data and applying the fitted model on a test data.
The BMARS based emulator method is very flexible since the basis functions are not predetermined and are adaptively chosen by the data. A Bayesian hierarchical model is used to obtain the posterior distribution of the unknown model parameters given the output data and a limited number of observations on the spatial field in full resolution. A hybrid sampling method, which is a combination of reversible jump MCMC method, Metropolis-Hastings method, and Gibbs sampling method, is used to sample from the posterior.
 
Spatial inverse problems have applications in the areas of groundwater flow, weather forecasting, chemical kinetics, reservoir characterization, and many other fields. For definiteness, we consider applications in the area of reservoir characterization where the single most influential input is the unknown permeability spatial field. The other model input parameters, such as porosity and the pore volume injected are considered to be known inputs. The output observations are fractional flow or water-cut data which is the fraction of water produced in relation to the total production rate in a two-phase oil-water flow reservoir. Furthermore, the ``fine-scale" (full-resolution or fine-grid) permeability is known at few well locations and the permeability data in a ``coarse-scale" (low-resolution or coarse-grid) are available from seismic traces. Some independent simulation runs of the computer forward simulator at the chosen design points are also performed offline to build the emulator. Evidently, our interest lies in assimilating the information obtained from simulated as well as the real data to infer about the unknown fine-scale permeability spatial field over a large domain. The Bayesian approach allows us to incorporate all these data from different sources in a hierarchical model setup.

The numerical results illustrate that the proposed BMARS model based on the transformed DCT coefficient can predict the output from the simulated test data adequately. The Bayesian model can also infer about the unknown spatial field together with its uncertainties. As expected, the computational efficiency in terms of CPU time for the emulator-based method is found to be very high when compared to the direct forward simulator-based methods.

The paper is organized as follows. In the next section, we discuss the DCT parametrization of the spatial field. In section 3, we formulate the Bayesian hierarchical model including the BMARS emulator. Section 4 details the Monte Carlo sampling procedure used to sample from the posterior. Section 5 presents the numerical results for a simulation study as well as a real application from a hydro-carbon reservoir.

\section{Parametrization of the spatial field using discrete cosine transform (DCT)}
The unknown calibration input in the emulator-based model is a high-dimensional spatial field. We consider an inexpensive form of parametrization of the fine-scale spatial field based on discrete cosine transform (DCT) (see \cite{ahmed}). The transformed DCT coefficients are then used as regressors in our Bayesian multivariate adaptive regression spline (BMARS) emulator. The transformation kernels used in the DCT are real cosine functions that are not dependent on the covariance structure of the spatial field. Using a Fast Fourier Transform (FFT) (see \cite{brig}), the DCT can be computed in $O(N log_2N)$ operations. This is much more computationally efficient than the Karhunen-Loeve (K-L) transformation, requiring a singular value decomposition of the covariance matrix as is of order $O(N^3)$ (see \cite{jain}, \cite{nara}). Since the DCT basis vectors are pre-determined and data-independent, they only need to be computed and stored once. The orthogonality of the DCT basis functions facilitates the computation of the inverse transform. Furthermore, the transformation is separable, hence a two-dimensional spatial field can be processed with one dimension at a time, resulting in a substantial reduction in computation time (see \cite{gong}, \cite{raoy}).
Using DCT transformation a one-dimensional field $Y(s)$ of $N$ grid points can be written as
\beq
\label{dct3}
Y(s)= \sum_{k=0}^{N-1}\alpha_k\theta_k cos\left[ \frac{\pi(2k+1)s}{2N}\right], 0\leq s \leq N-1,
\eeq
where,
\beq
\label{dct1}
\theta_k=\alpha_k\sum_{s=0}^{N-1}Y(s)cos\left[ \frac{\pi(2k+1)s}{2N}\right], 0\leq k \leq N-1,
\eeq
\beq
\label{dct2}
\alpha_k=\sqrt{\frac{2}{N}} \ \text{for} \ k=1,2, \ldots N-1, \ \alpha_0=\sqrt{\frac{1}{N}}.
\eeq
Here $\theta_k$'s are called transformed DCT coefficients.
Extensions of \ref{dct3} to higher dimensions can be  done by using the separability property of DCT, i.e., by applying the one-dimensional transform in each direction.
Using DCT transformation a two dimensional spatial field $Y(s)$, where $s=(s_x,s_y)$, of $N\times N$ grid points, can be written as
\beq
\label{dct6}
Y(s_x,s_y)= \sum_{i=0}^{N-1}\sum_{j=0}^{N-1}\alpha_i\alpha_j\theta_{ij}cos\left[ \frac{\pi(2s_x+1)i}{2N}\right]cos\left[ \frac{\pi(2s_y+1)j}{2N}\right] 
\eeq
where,

\beq
\label{dct5}
\theta_{ij}=\alpha_i\alpha_j\sum_{s_x}\sum_{s_y}Y(s_x,s_y).cos\left[ \frac{\pi(2s_x+1)i}{2N}\right]cos\left[ \frac{\pi(2s_y+1)j}{2N}\right], 0\leq i,j \leq N-1.
\eeq

\begin{figure}[tbp]
\centering
\includegraphics[width=6in,height=3.5in]{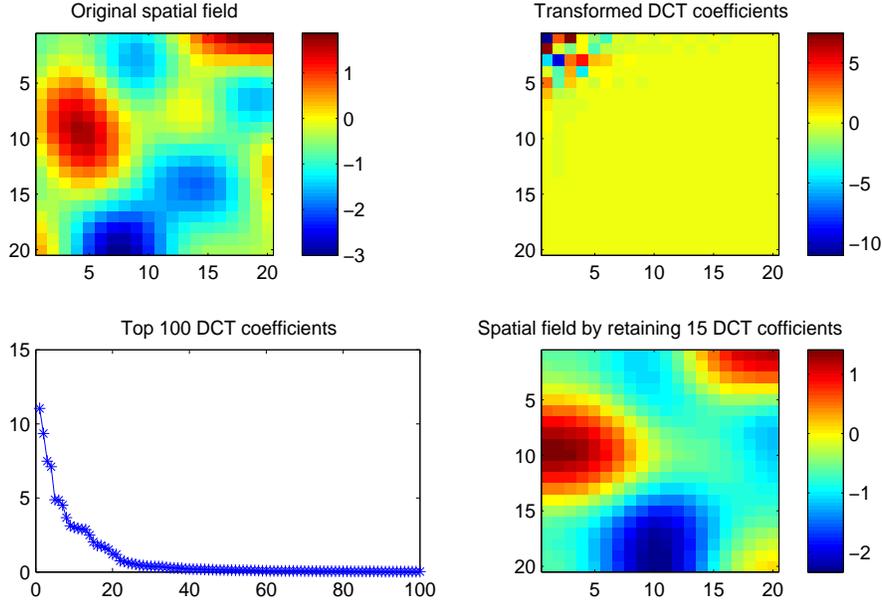}
\caption{A spatial field, the corresponding DCT coefficients and the spatial field obtained by the inverse DCT transform retaining the first 15 DCT coefficients.}
\label{dctfig1}
\end{figure}
 
Moreover, the DCT can be truncated to a few DCT coefficients while still preserving the prominent features and spatial dependency of the spatial field (see Figure \ref{dctfig1}). Generally, the concentration of the large DCT coefficients is on the top left corner of the transformed space. This clustering of coefficients corresponds to the modes with large-scale variations in the horizontal, vertical, and diagonal directions. When quantitative prior information is available, proper coefficients can be selected more systematically. For example, if the dominant features of the spatial field are known a priori to be vertically oriented then more coefficients should be selected from the left side of the DCT coefficients array. If the dominant features are horizontally oriented then more coefficients should be selected from the top of the DCT coefficient array. In a spatial inverse problem the spatial field is mostly unknown, so it is not possible to identify the retained modes by ordering the coefficients. In such cases, where prior information is unavailable, a reasonable alternative is to retain modes associated with coefficients inside a symmetric triangle or a square on the top left corner. 
Hence, applying truncated DCT in a regular grid, the spatial process $Y(s,\omega)$, where $\omega \in \Omega, s=(s_x,s_y) \in D \subset R^2$, can be written as
\beq
\label{dctp}
Y(s,w)=B_c(s)\bm{\theta}(w)
\eeq
where $\bm{\theta}$ is the vector of the truncated DCT coefficients and $B_c$ is the matrix of the corresponding DCT basis functions. 
Generally, the DCT coefficients decay very fast (see Figure \ref{dctfig1}) and we only need to retain a few DCT coefficients. Hence the dimension of $\bm{\theta}$ is usually very small compared to the dimension of the spatial field.
For example in Figure \ref{dctfig1}, a good approximation of the $20 \times 20$ dimension spatial field is obtained by retaining the first $15$ DCT coefficients. Given the truncated DCT coefficients $\bm{\theta}$, the spatial field can be obtained by the corresponding inverse DCT transformation as given in \ref{dctp}.
\par

\section{The Bayesian hierarchical model}
\label{sec_3}

In this section, we elaborate on the Bayesian hierarchical model used to construct the posterior distribution of the unknown model parameters given the observed output data. 
We consider two types of inputs: the first type is the unknown spatial field which we want to infer about using the output observations, the second type of inputs comprises all other model parameters which are completely known. 
In this article, we consider the uncertainty accounted for by the unknown input spatial field together with the uncertainties related to the computer codes. We first build an emulator which is an approximation to the unknown function that maps the inputs to the outputs, using both simulated and observed output data. 
Suppose we have $n_s$ run of the computer model or the forward simulator and the outputs are denoted as $z^{s}_{i},\ i=1, 2, \ldots n_s$. The corresponding known input model parameters are denoted as $x^{s}_{i},\ i=1,2, \ldots n_s$. The corresponding calibration inputs for the spatial field are generated by a Latin hypercube sampling of the DCT coefficients (as described in detail in subsection 3.5). Let us denote the simulated calibration inputs as $x^{t}_i,\ i=1,2, \ldots n_s$, and the corresponding input spatial field is obtained by the inverse DCT transformation. The statistical model for the simulated data is then given by
\beq
\label{emus}
z^{s}_{i}=\eta(x^{s}_{i},x^{t}_i)+\epsilon^{s}_{i}, \ i=1,2, \ldots, n_s.
\eeq
We also have additional $n_r$ \ $(n_r<<n_s)$ observed data from the output of the physical process given by $z^{r}_{i}, \ i=1,2 \ldots, n_r$. The corresponding known input variables are denoted as $x^{r}_{i}, \ i=1,2, \ldots n_r$ and the corresponding unknown spatial field is parametrized by the unknown DCT coefficients $\theta$.
The statistical model for the observed data is given by
\beq
\label{emur}
z^{r}_{i}=\eta(x^{r}_{i},\bm{\theta})+\epsilon^{r}_{i}, \ i=1,2, \ldots n_r.
\eeq
Here $\eta()$ denotes the statistical emulator based on multivariate adaptive regression splines, which is described in more detail in section 3.1. 

We assume that the emulator model errors for the forward simulator or computer code, $\epsilon^{s}_{i}$'s  are i.i.d. $N(0,\sigma_{z}^{2})$. The combined BMARS model errors and the observational errors for the output, $\epsilon^{r}_{i}$'s, are assumed to be i.i.d. $N(0,\tau_{z}\sigma_{z}^{2})$. For the observed data, the scale multiplier for the error variance, $\tau_{z}$, is considered in order to incorporate the model discrepancy factor. The model discrepancy factor signifies the difference in the true mean output of the real-world physical process and the corresponding computer model output for the given inputs.

Note that this model discrepancy factor representation is different from the usual model calibration approach \citep{ken}. In most of the literature, a separate model (e.g. a GP) is assumed for the model discrepancy. On the other hand, \cite{JB14} showed that without informative priors on the model discrepancy parameters the model discrepancy error is confounded with the other types of errors. So without any prior information on the model discrepancy error, we propose a simple alternative for our application, where the variance of the combined discrepancy error, emulator error, and observational error are considered as a multiple of the emulator model error variance. This formulation also allows us to use a Gibbs step while updating $\tau$ in the MCMC method. This is described in more detail in section 4.

In addition to the output observations, we also assume that a ``coarse-scale" or ``low resolution" data is available for the unknown spatial field, denoted by $\bm{y}_c$. A very few ``fine-scale" or full-resolution observations are also available for the spatial field denoted by $\bm{y_{o}}$. We would like to integrate all these available data ($\bm{z^r}$, $\bm{y}_c$ and $\bm{y}_o$) in the Bayesian hierarchical model to quantify the uncertainty in the input spatial field.

The hierarchical model for the posterior distribution of the model parameters is given by 
\ber
\label{ps1}
P(\bm{\theta}, \sigma_{z}^{2}, \tau_{z}|\bm{x^{t}}, \bm{x^{r}}, \bm{x^{s}}, \bm{z^{r}}, \bm{z^{s}}, \bm{y}_c, \bm{y}_o) &\propto& P(\bm{\theta}, \sigma_{z}^{2}, \tau_{z},  \bm{z^{r}}, \bm{z^{s}}, \bm{y}_c, \bm{y}_o|\bm{x^{r}}, \bm{x^{s}}, \bm{x^{t}}) \nonumber \\
&=& P(\bm{z^{r}}, \bm{z^{s}}|\bm{\theta}, \sigma_{z}^{2}, \tau_{z}, \bm{x^{r}}, \bm{x^{s}}, \bm{x^{t}}) \nonumber \\
&\times& P(\bm{y}_c|\bm{\theta})P(\bm{y}_o|\bm{\theta})P(\bm{\theta}, \tau_{z}, \sigma_{z}^{2}).
\eer
The first term on the second line of the right-hand side of the hierarchical model (\ref{ps1}) is the likelihood of the output data given the inputs and model parameters. Note that the likelihood contains the unknown function $\eta$, which maps the inputs to the outputs.
The second term, $P(\bm{y}_c|\bm{\theta})$, denotes the probability distribution of the coarse-scale data  given the spatial field. The third term, $P(\bm{y}_o|\bm{\theta})$, denotes the probability distribution of the observed fine-scale spatial data given the DCT coefficients. The last term, $P(\bm{\theta}, \tau_{z}, \sigma_{z}^{2})$, denotes the prior distribution for the model parameters.
In the following subsections, we will first elaborate on how we model each of the terms on the right-hand side of the Bayesian hierarchical model (\ref{ps1}). Then we will discuss the design techniques used for sampling the inputs for the $n_s$ simulation runs.

\subsection{Modeling the likelihood using Bayesian MARS emulators}
\label{sec_4}

In this section, we describe the Bayesian  multivariate adaptive regression spline(BMARS) approximation of the forward simulator as described in \ref{emus} and \ref{emur}. We assume that there are $k_1$ input variables that are completely known for each output, so that
$\bm{x}^{s}_{i}=(x^{s}_{i1}, x^{s}_{i2},\ldots, x^{s}_{ik_1}), \ i=1,2, \ldots n_s$ and $\bm{x}^{r}_{i}=(x^{r}_{i1}, x^{r}_{i2}, \ldots, x^{r}_{ik_1}), \ i=1,2, \ldots n_r$. Let $X_s$ and $X_r$ denote the corresponding matrices obtained by augmenting the vectors $\bm{x}^{s}_{i}$'s and $\bm{x}^{r}_{i}$'s respectively.
Also, it is assumed that $k_2$ coefficients are retained in the DCT transformation of the spatial field, i.e., $\bm{x}^{t}_i=(x^{t}_{i1}, x^{t}_{i2}, \ldots, x^{t}_{ik_2}), i=1,2, \ldots n_s$. Let $X_t$ be the matrix obtained by augmenting these $\bm{x}^{t}_i$'s. Similarly, let $X_\theta$ be the matrix obtained by augmenting $\theta$'s $n_r$ times. Combining the simulated data and the observed data we can write the model \ref{emus} and \ref{emur} as
\beq
\label{emucomb}
Z=\eta(X)+ \bm{\varepsilon}
\eeq
where, 
$X = \left( \begin{array}{cc}
 X_s & X_t \\
X_r & X_{\theta}
\end{array} \right),\
Z=\left( \begin{array}{c}
\bm{z^s}\\
\bm{z^r}
\end{array} \right)
= \left( z^{s}_{1},z^{s}_{2}, \ldots, z^{s}_{n_s},z^{r}_{1},z^{r}_{2}, \ldots, z^{r}_{n_r} \right)'$
and
$\bm{\varepsilon}= \left( \begin{array}{c}
\bm{\epsilon^{s}} \\
\bm{\epsilon^{r}}
\end{array} \right) = \left( \epsilon^{s}_{1},\epsilon^{s}_{2} \ldots \epsilon^{s}_{n_s},\epsilon^{r}_{1} \ldots \epsilon^{r}_{n_r} \right)'$.
The multivariate adaptive regression splines (MARS) (see \cite{fried}) is a regression method where the unknown simulator or ``black box" is approximated by a regression function as given below:
\beq
\label{bmars1}
\eta(x)=\sum_{i=1}^{m}\beta_iB_i(x)
\eeq
where, the basis function $B_i(x)$ is given by
\beq
B_i(x)=\left\{
\begin{array}{l l}
1 & i=1  \\
 \prod_{j=1}^{J_i}\left[s_{ji}.(x_{\nu(ji)}-t_{ji})\right]_{+}, & i=2,3,..\end{array} \right. 
\eeq
Here $(.)_{+}=max(0,.)$ and $J_i$ is the degree of the interaction of $B_i$. The $s_{ji}$, which we shall call the sign indicators,
equal $\pm 1$ and the $\nu(j,i)$'s are the index of the predictor variable
which is being split on the $t_{ji}$ (known as the knot points). The knot points give
the position of the splits. The $\nu(j,.)\ (j=1\ldots J)$ are constrained to be distinct so that each predictor only appears once
in each interaction term. Frequently some
maximum order of interaction $I$ is assigned to the model
such that $J_i\leq I (i = 1\ldots m)$. We let $T_i \in \{1,2, \ldots N_I\} (N_I=\sum_{i=1}^{I} {{k1+k2}\choose{i}})$ to denote
the type of basis function $B_i$. Thus $T_i$, in effect, just tells us
which predictor variables we are splitting on, i.e. what the
values of $\nu(1,i),\ldots  ,\nu(J_i,i)$ are.
We denote all the parameters in the model as $(m,\bm{c}^m, \bm{\beta^m})$. Here $\bm{c}^m=(\mathcal{V}_1, \ldots,\mathcal{V}_m)$ where each $\mathcal{V}_i$ is the 
$(1 +2J_i)$ dimensional vector
$(T_i,t_{1i}, s_{1i},\ldots, t_{Ji}, s_{Ji})$ which corresponds to the basis
function $B_i$.
After including the BMARS parameters in the hierarchical model, the posterior distribution (\ref{ps1}) is modified to the following form.
\ber
\label{ps2}
\pi(\bm{\theta},\sigma_{z}^{2},\tau_{z}, m, \bm{\beta^m}, \bm{c}^m)&=&P(\bm{\theta},\sigma_{z}^{2},\tau_{z}, m, \bm{\beta^m}, \bm{c}^m|X_t, X_r,X_s,Z,\bm{y}_c,\bm{y}_o) \nonumber \\
&\propto& P(\bm{\theta},\sigma_{z}^{2},\tau_{z},m,\bm{\beta^m},\bm{c}^m, Z, \bm{y}_c, \bm{y}_o|X_t, X_r,X_s) \nonumber \\
&=&P(Z|X, m, \bm{\beta^m}, \bm{c}^m,\sigma_{z}^{2},\tau_{z})P(\bm{y}_c|\bm{\theta})P(\bm{y}_o|\bm{\theta}) \nonumber \\
&\times& P(\bm{\theta})P(\tau_{z})P(\sigma_{z}^{2})P(\bm{\beta^m}|m)P(\bm{c}^m|m)P(m).
\eer
We assume
\begin{equation}
\epsilon^{s}_{i} \iid N(0, \sigma_{z}^{2}), i=1,2, \ldots, n_s,
\end{equation}
and
\begin{equation}
\epsilon^{r}_{i} \iid N(0, \tau_{z}\sigma_{z}^{2}), i=1, 2, \ldots, n_r.
\end{equation}
Hence, the likelihood $P(Z|X, m, \bm{c}^m, \bm{\beta^m}, \sigma^2_z, \tau_z)$ is the pdf of the  multivariate Normal distribution with mean $\sum_{i=1}^{m}\beta_i B_i(X)$  and
variance covariance matrix $\Sigma=\sigma_{z}^2 S$, where $S = \left( \begin{array}{cc}
I_{n_s}& \mathbf{0}\\
\mathbf{0}& \tau_{z}I_{n_r}
\end{array} \right) $.

\subsection{Modeling the coarse-scale data}
\label{sec_5}
In many cases, low resolution or ``coarse-scale" data are readily available which could be easily incorporated into the hierarchical model to reduce the uncertainty of the fine-scale spatial field. To include the coarse-scale data we would use an ``upscaling" procedure linking the coarse and the fine-scale data.
The simplest way to describe the upscaling procedure in the spatial domain is the use of
spatial block averages of the fine-scale spatial data to obtain the coarse-scale data. We need to modify this averaging idea in such a way that the average output obtained by solving the forward model (and the corresponding boundary conditions) in this upscaled spatial field is very close to the output obtained by solving the forward model in the fine scale.
The main theme of the procedure is that given a fine-scale spatial field $Y$, we can use an upscaling operator $L$ (it can be averaging or more complicated integrations with boundary conditions), such that that the coarse-scale data $\bm{y}_c$ can be expressed as $\bm{y}_c=L(Y)+\epsilon_{c}$. Here $\epsilon_{c}$ is the random error that explains the variations of the observed coarse-scale data from the deterministic upscaling procedure. As we have parameterized the spatial field $Y$ using the DCT transformation the corresponding model in terms of the DCT coefficients is given by

\begin{equation}
\bm{y}_c=L_c(\bm{\theta})+\bm{\epsilon_c} 
\end{equation}
where, $L_c$ can be looked upon as an operator whose input is the transformed DCT coefficients of the fine-scale spatial field and output is the coarse-scale spatial field. We assume that the error $\bm{\epsilon_c}$ follows a multivariate Normal distribution with mean $\bm{0}$ and covariance matrix $\sigma_{c}^{2}I$, i.e., $\bm{y}_c|\bm{\theta},\sigma_{c}^{2} \sim MVN(L_c(\bm{\theta}),\sigma_{c}^{2}I)$. The prior distribution of $\sigma_{c}^{2}$ is assumed to be Inverse Gamma$(a_c, b_c)$. Furthermore, after integrating out $\sigma_{c}^{2}$ the marginal distribution of $\bm{y}_c$ given $\bm{\bm{\theta}}$ is given by
\begin{equation}\label{corslikeemu}
P(\bm{y}_c|\bm{\theta})\propto \frac{\Gamma(a_c+N^*/2)}{\bigl[b_c+\frac{1}{2}\bigl(\bm{y}_c-L_c(\bm{\theta})\bigr)'\bigl(\bm{y}_c-L_c(\bm{\theta})\bigr)\bigr]^{(a_c+N^*/2)}}.
\end{equation}
where $N^*$ is the number of observed data on the coarse-scale spatial field.

\subsection{Modeling the observed fine-scale data}
\label{sec_6}

As mentioned before, the fine-scale observations are also obtained at a few locations of the spatial field. The model for such observed fine-scale spatial data, $\bm{y}_o$, is given by
\begin{equation}
\bm{y}_o=\bm{y_{p}}+ \bm{\epsilon_{k}} 
\end{equation}
where, $\bm{y_{p}}$ is the the fine-scale spatial field at the given locations of $\bm{y_{o}}$ obtained from the inverse DCT transformation of $\bm{\theta}$ as described before. $\bm{\epsilon_{k}}$ is the model error for the DCT approximation. We assume $\bm{\epsilon}_{k}$ follows a multivariate Normal distribution with mean $0$ and covariance matrix $\sigma_{k}^2I$, i.e., $\bm{y}_o|\bm{\theta}, \sigma_{k}^2 \sim MVN(\bm{y_{p}},\sigma_{k}^2I)$. The prior for $\sigma_{k}^{2}$ is assumed to be Inverse Gamma$(a_k, b_k)$. After integrating out $\sigma_{k}^{2}$ the conditional distribution of $\bm{y}_o$ given $\bm{\theta}$ is given by
\begin{equation}\label{obslike}
P(\bm{y}_o|\bm{\theta})\propto \frac{\Gamma(a_k+N_{obs}/2)}{[b_k+\frac{1}{2}(\bm{y}_o-\bm{y_{p}})'(\bm{y}_o-\bm{y_{p}})]^{(a_k+N_{obs}/2)}}.
\end{equation}
where $N_{obs}$ is the number of observations on the fine-scale spatial field.

\subsection{Prior specification}
\label{sec_7}
Bayesian multivariate adaptive regression spline assigns a prior distribution to every unknown parameter in the model.
We assume vague, but proper, prior for $\sigma^2_z$, where
$\sigma_z^2 \sim \text{Inverse \ Gamma}(a_z,b_z)$. Similarly we assume $\tau_z \sim \text{Inverse \ Gamma}(a_{\tau},b_{\tau})$. The prior for $T_i$ are assumed to be uniformly distributed on $\{1,2, \ldots N\}$. The prior for sign indicators, $s_{ji}$, and knot points, $t_{ji}$, are also assumed to be uniformly distributed on the sets $\{1,-1\}$ and $\{1,2, \ldots n\}$ respectively. We use another vague, but proper prior, for the coefficients of the basis functions, i.e., we assume the $\beta_i \sim N(0, \alpha\sigma_{z}^2)$, where $\alpha$ is very large. The prior distribution of $m$ is taken to be a truncated Poisson distribution with parameter $\lambda$, truncated at $m_{max}$.
The prior distribution for $\bm{\theta}$ is considered as following.\\
$\bm{\theta}|\sigma_{\theta}^{2} \sim MVN(0,\sigma_{\theta}^{2}I)$ and  
$\sigma_{\theta}^{2} \sim  \text{Inverse \ Gamma}(a_o, b_o)$.
After integrating out $\sigma_{\theta}^{2}$ we obtain the marginal prior distribution as
\begin{equation}\label{prtheta}
P(\bm{\theta}) \propto \frac{\Gamma(a_o+k_2/2)}{[b_o+\frac{1}{2}\bm{\theta}'\bm{\theta}]^{(a_o+k_2/2)}}.
\end{equation}

\subsection{Design of the simulation experiments}

OOne of the very important issues in implementing the emulator model is the choice of input configurations at which the simulator model is executed to obtain the training data. The main objective of choosing the input design configuration is to learn how the unknown function maps the input to the output over well-spaced input points that cover the wide region of interest as close as possible. In our problem, the unknown inputs of the forward model is a spatial field. The usual way to create multiple realizations of spatial random field is to draw a simple random sample from a Gaussian process with a proper correlation structure. Since the computer simulator is very expensive for a complicated physical process, the number of samples of the spatial field has to be kept small for practical reasons. In that case, a more accurate assessment of the emulator can be obtained when a more efficient sampling method such as Latin hypercube sampling (\cite{mack}, \cite{ross}) is used. 

In this article, we use the following design scheme for our emulator. As discussed before, very few realizations of the fine-scale spatial field $y_o$ are available, but the data from a relatively coarser scale, $\bm{y}_c$ is available. We use these coarse-scale data for the design purpose. In this situation, the available coarse-scale data is transformed to fine-scale data by replacing every element in the fine-scale of a coarse block with the corresponding coarse-scale value. Then DCT transformation is applied to such fine-scale data and the DCT coefficients are truncated up to a desired degree of accuracy. Suppose the transformed DCT coefficients are $\bm{\theta}_{obs}$. A large number of samples of the DCT coefficients, say $N_s$ samples, are then obtained using Latin hypercube sampling from a multivariate Normal distribution with mean $\bm{\theta}_{obs}$ and variance $\gamma \mathbf{I}$.  
Such Latin Hypercube samples can be obtained by slightly shifting the simple random samples obtained from the target multivariate distribution (see \cite{stein, mondal20}).
Then $n_s$ samples, having the minimum distance between the set of points, are selected from the original $N_s$ LHS samples. The corresponding $n_s$ input spatial fields are obtained from the inverse transformation of the sampled DCT coefficients.

\section{Sampling from the posterior}

The Bayesian model casts the inverse solution as a posterior probability distribution over the model parameters. 
Here, we use a hybrid sampling method, which is a combination of Gibbs sampling and Metropolis-Hastings algorithm, to sample from the joint posterior distribution, $\pi(\bm{\theta},\sigma_{z}^{2},\tau_{z}, m, \mathbf{\beta^m}, \bm{c}^m)$. Note that the number of basis functions in the BMARS model is not fixed, hence the dimension of the parameter space of the posterior is also random. We use the reversible jump MCMC algorithm, as introduced by \cite{rjmc95}, to sample from the parameter space of varying dimensions. An algorithm of the sampling procedure from the posterior distribution is given by \ref{alms}.

\begin{algorithm}
\caption{Hybrid Sampling Algorithm}
\label{alms}
{\bf{Step 1.}} First we would like to sample from the joint conditional distribution of $m, \bm{c}^m,\bm{\beta^m},\sigma_{z}^{2}$ given $\bm{\theta},\tau_z$. In order to sample from this joint distribution we do the following three sub-steps.
\ben
\item
The marginal distribution of $m, \bm{c}^m$ given $\bm{\theta},\tau_z$ can be computed up to a constant of proportionality by integrating over $\sigma_{z}^{2}$ and $\bm{\beta}$. We sample from this marginal distribution using reversible jump MCMC technique where at each step we use one of the following types of moves: (a)the addition of a basis function (BIRTH); (b) deletion of a basis function (DEATH); (c) a change in a  knot location (CHANGE). The BIRTH move is carried out by choosing uniformly a type of basis function, say $T_i$ to include in the model. Then a knot location and sign indicator for each of the $J_i$ factors in the new basis is chosen uniformly. In the DEATH move, the deletion of a basis function is done in such a way as to make the jump reversible. This is done by removing a uniformly chosen basis function from the present set of basis functions (except the constant basis function).
\item The marginal distribution of $\bm{\beta}$ given $m, c^m,\tau_z,\bm{\theta}$ can be expressed in a closed-form after integrating over $\sigma_{z}^{2}$, so a Gibbs sampling step is used to sample from this marginal conditional distribution.
\item The conditional distribution of $\sigma_{z}^{2}$ given $m, \bm{c}^m,\bm{\beta^m}, \tau_z,\bm{\theta}$ can be expressed in a closed form, so a Gibbs sampling step is used to sample from this conditional distribution.
\een
{\bf{Step 2.}} The conditional distribution of $\tau_z$ given $m, \bm{c}^m,\mathbf{\beta^m}, \sigma_{z}^{2},\bm{\theta}$ can be expressed in a closed form, so a Gibbs sampling step is used to sample from this conditional distribution.
\\
{\bf{Step 3.}} The conditional distribution of $\bm{\theta}$ given $m, \bm{\beta^m}, \bm{c}^m,\sigma_{z}^{2},\tau_z$ can be expressed up to a constant of proportionality, so a Metropolis-Hastings step is used to sample from this conditional distribution. 
\end{algorithm}

The details about the conditional and marginal distributions used in Algorithm \ref{alms} are given below.
{\bf{Step 1.}}
\begin{eqnarray}
\label{mrglk}
P(m, \bm{c}^m|\bm{\theta},\tau_z,  X_t, X_r,X_s,Z, \bm{y}_c, \bm{y}_o)&\propto& \int_{\beta} \int_{\sigma_{z}^2}P(Z|X,\sigma_{z}^2, \tau_z,m,\bm{c}^m,\mathbf{\beta^m}) \nonumber \\
&\times&P(\bm{c}^m |m)P(\beta|m)P(m)P(\sigma_{z}^2)d\sigma_{z}^2 d\beta \nonumber \\
&=&\pi_1(m, \bm{c}^m|\bm{\theta},\tau_z), \text{(say)}
\end{eqnarray}
It can be shown that 
\beq
log(\pi_1)=C_1-(\frac{n}{2}+a_z)log(d)-\frac{m}{2}\tau_z +(m-1)log(\lambda)-log(p!)-\frac{m}{2}log(\alpha)-log(N_I)-\sum_{j=1}^{m}J_j log(2n),
\eeq
where, 
\ber
d=2b_z+Z'S^{-1}Z+(\mathbf{B}'S^{-1}Z)'\Sigma_k^{-1}(\mathbf{B}'S^{-1}Z),\ \Sigma_k=\left[ \mathbf{B}'S^{-1}\mathbf{B}+I/\alpha\right], \nonumber \\
p=k_1+k_2, \ C_1=\frac{1}{2}log|\Sigma_k^{-1}| \ \text{and} \ \mathbf{B}=(B_1,B_2,\ldots, B_m).
\eer
Hence, the acceptance probability of reversible jump MCMC step as described in the first part of step $1$ of the algorithm \ref{alms} can be written as
\beq
\label{rjal1}
\alpha=min \left( 1, \frac{\pi_{1}(m',\bm{c}'^{m'}|\tau_z,\bm{\theta})S((m',\bm{c}'^{m'})\rightarrow(m,\bm{c}^{m}))}{\pi_{1}(m,\bm{c}^{m}|\tau_z,\bm{\theta})S((m,\bm{c}^{m})\rightarrow(m', \bm{c}'^{m'}))}\right)
\eeq
where,$(m,\bm{c}^{m})$ denotes the current model parameters and $(m',\bm{c}'^{m'})$ denotes the proposed model parameters. At each iteration we either have a ``birth" move or a ``death" move or a ``change" move step with probability $b_m$, $d_m$ and $\rho_m$ respectively. The proposal ratio for a ``birth" move is given by
\ber
\frac{S((m',\bm{c}'^{m'})\rightarrow(m,\bm{c}^{m}))}{S((m,\bm{c}^{m})\rightarrow(m,\bm{c}'^{m'}))}&=&\frac{P\left(\text{propose death} (m+1,\bm{c}'^{m+1}) \rightarrow(m,\bm{c}^{m})\right)}{P\left(\text{propose birth}(m,\bm{c}^{m})\rightarrow(m+1,\bm{c}'^{m+1})\right)} \nonumber \\
&=& \frac{d_{m+1}/m}{b_m/[N_I(2n)^{J_{m+1}}]},
\eer

The proposal ratio for a ``death" move is given by
\ber
\frac{S((m',\bm{c}^{m'})\rightarrow(m,\bm{c}^{m}))}{S((m,\bm{c}^{m})\rightarrow(m,\bm{c}'^{m'}))}&=&\frac{P\left(\text{propose birth} (m-1,\bm{c}'^{m-1}) \rightarrow(m,\bm{c}^{m})\right)}{P\left(\text{propose death}(m,\bm{c}^{m})\rightarrow(m-1,\bm{c}'^{m-1})\right)} \nonumber \\
&=& \frac{b_{m-1}/[N_I(2n)^{J_{m-1}}]}{d_m/(m-1)}.
\eer

The proposal ratio for a ``change" move is always $1$.  Note that here $b_m+d_m+\rho_m=1, \ \forall m$. In particular we take $b_m=d_m=\rho_m=\frac{1}{3} \ \forall m=2,3, \ldots m_{max}$, \ $b_1=1,d_1=\rho_1=0$ \ and \ $b_{m_{max}}=0,d_{m_{max}}=1,\rho_{m_{max}}=0$

Similarly, the Gibbs step in the second part of step $1$ of the algorithm is carried out by generating samples from the marginal conditional distribution of $\beta^m$given $m, \bm{c}^m,\tau_z,\bm{\theta}, X_r, X_t, X_s, Z,\bm{y}_c, \bm{y}_o$ given by
\beq
\beta^m|m, \bm{c}^m,\tau_z,\bm{\theta}, X_t, X_r, X_s, Z, \bm{y}_c, \bm{y}_o \sim t_{n+2a_z}\Bigr(\Sigma_k[\mathbf{B}'\Sigma^{-1}Z], \frac{d}{n+2a_z}\Sigma_k\Bigl), 
\eeq

In the third part of step $1$ of the algorithm, the Gibbs step is done by generating random samples from the conditional distribution of 
$\sigma_{z}^{2}$ given $m, \bm{c}^m,\mathbf{\beta^m}, \tau_z,\bm{\bm{\theta}}, X_t, X_r,X_s,Z, \bm{y}_c, \bm{y}_o$, which is given by
\beq
\sigma_{z}^{2}|m, \bm{c}^m,\mathbf{\beta^m}, \tau_z,\bm{\bm{\theta}}, X_t, X_r,X_s,Z, \bm{y}_c, \bm{y}_o \sim \text{Inverse Gamma}(\delta_{z1},\delta_{z2})
\eeq
where,
\beq
\delta_{z1}=a_z+\frac{m+n}{2}, \delta_{z2}=b_z + \frac{(Z-\mathbf{B}\beta)'\Sigma^{-1}(Z-\mathbf{B}\beta)}{2}+\frac{\beta^2}{2\alpha^2}
\eeq
\noindent
{\bf{Step 2.}}
For a given $m,\bm{c}^m$ we write the matrix of the basis functions defined on BMARS as
$\mathbf{B}
= \left( \begin{array}{c} \mathbf{B_s} \\ \mathbf{B_r} \end{array} \right)$. The Gibbs sampling in step two of the algorithm \ref{alms} is carried out by sampling from the conditional distribution of $\tau_z$ given $m, \bm{c}^m,\bm{\beta^m}, \sigma_{z}^{2},\bm{\theta},X_t, X_r,X_s,Z, \bm{y}_c, \bm{y}_o$, which is given by
\beq
\tau_z|m, \bm{c}^m,\mathbf{\beta^m}, \sigma_{z}^{2},bm{\theta}, X_t, X_r, X_s, Z, \bm{y}_c, \bm{y}_o \sim \text{Inverse Gamma}(\delta_{\tau1}, \delta_{\tau2})
\eeq
where,
\beq
\delta_{\tau1}= \frac{n_r}{2} + a_{\tau},\delta_{\tau2}=\frac{(Z_r -  \mathbf{B_r}\beta^m)'(Z_r -  \mathbf{B_r}\beta^m)}{2\sigma_{z}^{2}} + b_{\tau}
\eeq
{\bf{Step 3.}}
The condition distribution of $\bm{\theta}$ given $m, \bm{c}^m,\mathbf{\beta^m}, \sigma_{z}^{2},\tau_z, X_t, X_r, X_s, Z, \bm{y}_c, \bm{y}_o$ as used in step three of the algorithm is described below. Let us denote \\
$P(\bm{\theta}|m, \bm{c}^m,\bm{\beta^m}, \sigma_{z}^{2},\tau_z, X_t, X_r,X_s,Z, \bm{y}_c, \bm{y}_o)$ as $\pi_2(\bm{\theta}|m, \bm{c}^m,\mathbf{\beta^m}, \sigma_{z}^{2},\tau_z)$, then
\ber
log(\pi_2)&\propto& -(Z_r -  \mathbf{B_r}\beta^m)'(Z_r - \mathbf{B_r}\beta^m)/(2\sigma_{z}^{2}\tau_z) \nonumber\\
&-& (a_c+N^*/2)log\left(b_c+(\bm{y}_c-L_c(\bm{\theta}))'(\bm{y}_c-L_c(\bm{\theta}))/2\right) \nonumber \\
&-&(a_k+N_{obs}/2)log\left(b_k+(\bm{y}_o-y_p)'(\bm{y}_o-y_p)/2\right) \nonumber \\
&-&(a_o+k_2/2)log\left(b_o+\bm{\theta}'\bm{\theta}/2\right).
\eer
The Metropolis-Hastings sampling in the third step of the algorithm is carried out by first proposing a new parameter $\bm{\theta}'=\bm{\theta}+h\xi$, where $\xi$ is a random variable and $h$ is the jump size. The acceptance probability of $\bm{\theta}'$ is given by
\beq
\alpha=min\left( 1, \frac{\pi_2(\bm{\theta}')q(\bm{\theta}|\bm{\theta}')}{\pi_2(\bm{\theta})q(\bm{\theta}'|\bm{\theta})}\right)
\eeq
where $q(\bm{\theta}|\bm{\theta}')$ is the proposal distribution of $\bm{\theta}$ given $\bm{\theta}'$.

\section {Simulation and real examples from reservoir models}

The spatial inverse problem has applications in many fields such as groundwater flow, chemical kinetics, weather forecasting, and reservoir characterization. For the definiteness, here we only consider examples from reservoir characterization. Petroleum reservoirs are complex geological formations that exhibit a wide range of physical and chemical heterogeneities. Geostatistics, and more specifically, stochastic modeling of reservoir heterogeneities are being increasingly considered by reservoir analysts and petroleum engineers for their potential in generating
more accurate reservoir models together with realistic measures of spatial uncertainty.
The goal of reservoir characterization is to provide a numerical model of reservoir attributes such as hydraulic conductivities (permeability), storativities (porosity), and fluid saturation. These attributes are then used as inputs to the complex forward model
represented by various flow simulators to forecast future reservoir performance
and oil recovery potential. In most flow situations, the single most influential input is the permeability spatial field, $k$ in our notation. Permeability is an important concept in porous media flow (such as the flow of the underground oil). Physically, permeability arises both from the existence of pores and from the average structure of the connectivity of pores. The
permeability is measured on a positive scale and its distribution is generally positively skewed, so we take logarithm transformation for our modeling convenience, i.e., $Y = log(k)$. The main available response is the fractional flow or the water-cut data (denoted by $z$), which is the fraction of water produced in relation to the total production rate in a two-phase oil-water flow reservoir. There are two kinds of inputs of the forward simulator: (i) a permeability spatial field $Y$ which is parametrized by DCT transformed coefficients $\bm{\theta}$ and (ii) the pore volume injected. Given the inputs of the model, the output or the water-cut observations can be obtained from Darcy's law and conservation of mass, which contains several partial differential equations. The goal of our inverse model is to learn about the unknown input spatial field given the output ($\bm{z_r}$), some coarse-scale realization of the permeability field ($\bm{y_c}$), and a few fine-scale realizations of the spatial field ($\bm{y_o}$) in the well locations.

\subsection{Numerical results for a simulated reservoir example} 

\begin{figure}[htbp!]
\centering
\includegraphics[width=4.5in,height=2in]{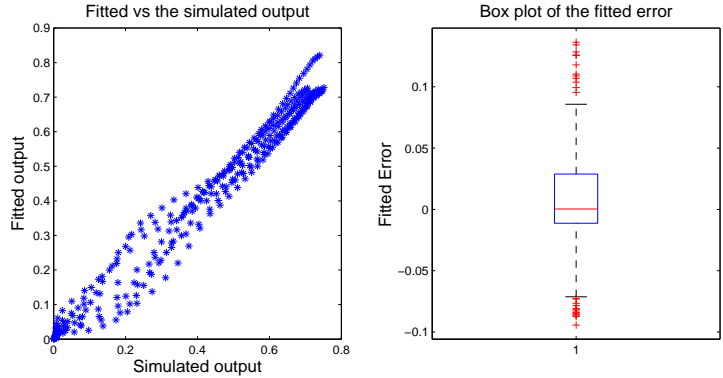}
\caption{Left: Cross plot of fitted vs simulated test data, Right: Box plot of the residuals for the test data}
\label{emufitted1}
\end{figure}

\begin{figure}[htbp!]
\centering
\includegraphics[width=3.5in,height=2.5in]{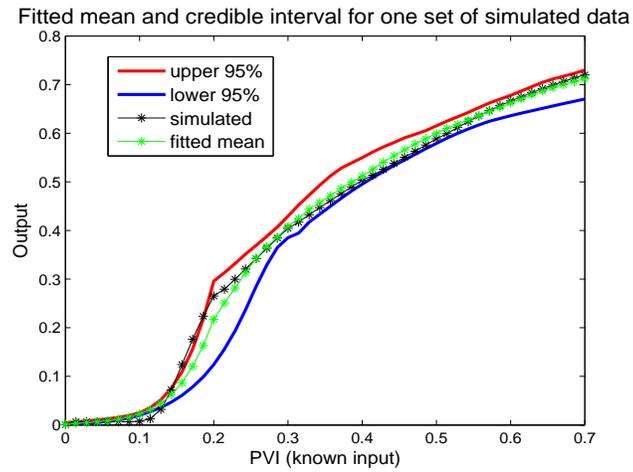}
\caption{Fitted mean and $95\%$ credible interval for one set of test data using the emulator}
\label{emufitted2}
\end{figure}

In our first example, we consider a simulated data from reservoir characterization. As discussed before, the output is the fraction flow or water-cut data and the two types of inputs are a $25\times25$ spatial field on a unit square and pore volume injected. Each grid for fine-scale data is of $0.04 \times 0.04$ square unit. First, we generate $100$ samples of $15$ DCT coefficients using Latin hypercube sampling from multivariate Normal distribution. Each of these sets of $15$ DCT coefficients corresponds to a log permeability field obtained by the respective inverse DCT transformation. Thus we have $100$ simulated realizations of the input spatial field. The other input variable is the pore volume injected (rescaled to $0-1$) at the injector wells. The simulated output is the fractional flow or water-cut data which is obtained by running the computer simulator of the forward model. For each simulated spatial field we have $50$ outputs corresponding to $50$ known input of second type (pore volume injected). An additional spatial permeability field is obtained using GEO-R software which is treated as a reference permeability field. We treat this spatial field as unknown which we want to calibrate. We use the computer forward model to simulate the corresponding outputs.  Since no real data is used in the model, the model discrepancy factor is not considered here, i.e.,  $\tau_z=1$. Before the calibration problem is solved, we first build an emulator based on a portion of the simulated data, called the training data, and see how our emulator performs on the test data. So we first divide the $5000$ simulated data set on the inputs and the outputs corresponding to the $100$ simulated spatial field into two parts. The first $4500$ data set corresponding to $90$ simulated spatial permeability field is called the training data set. The rest $500$ data set corresponding to $10$ permeability field is regarded as the test data set. First, we built the BMARS emulators where the regressors are the DCT coefficients of the permeability field and the pore volume injected. The response is the logit transformation of the water-cut data. Then we use the fitted model on the training data to predict the output for the test data.  This process is called computer model validation.  In Figure \ref{emufitted1} the scatter plot of the mean of fitted output versus the mean of the simulated output shows that all the observations lie almost on the straight line through the origin. Also, from the box plot of the predicted errors, we can see that the median of the errors is close to zero. The simulated output, the corresponding fitted mean, and the $95\%$ credible interval are shown in Figure \ref{emufitted2}. All the above-mentioned plots indicate that the BMARS emulators can predict the output very well.

\begin{figure}[h!]
\centering
\includegraphics[width=6in,height=4.5in]{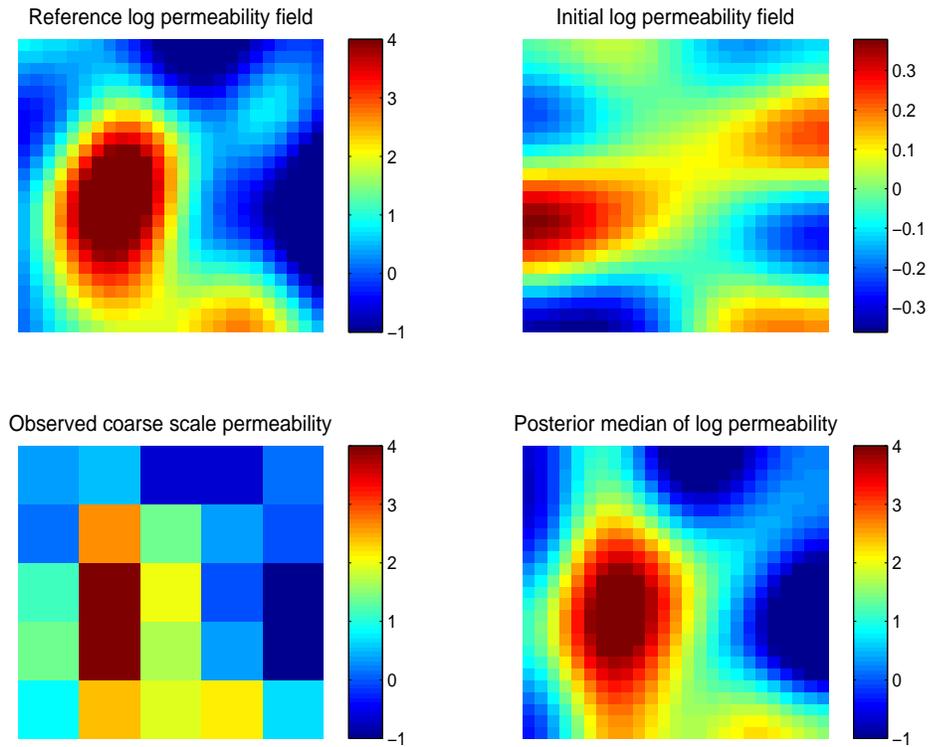}
\caption{Log permeability fields for the simulated example. Top left: Reference field, Top right: Initial field of the Markov chain, Bottom Left: Observed coarse-scale permeability, Bottom right: Posterior median.}
\label{emuperm2}
\end{figure}
\begin{figure}[h!]
\centering
\includegraphics[width=4in,height=3.5in]{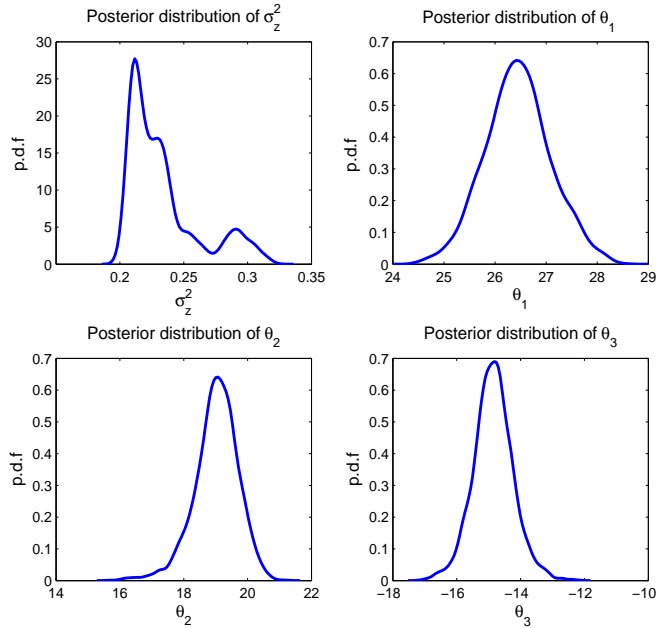}
\caption{Posterior distributions for the simulated example. Top Left: Posterior density of $\sigma_{z}^{2}$, Top Right: Posterior density of $\theta_1$, Bottom Left: Posterior density of $\theta_2$, Bottom Right: Posterior density of $\theta_3$.}
\label{simupdf1}
\end{figure}
\begin{figure}[h!]
\centering
\includegraphics[width=5.5in,height=4in]{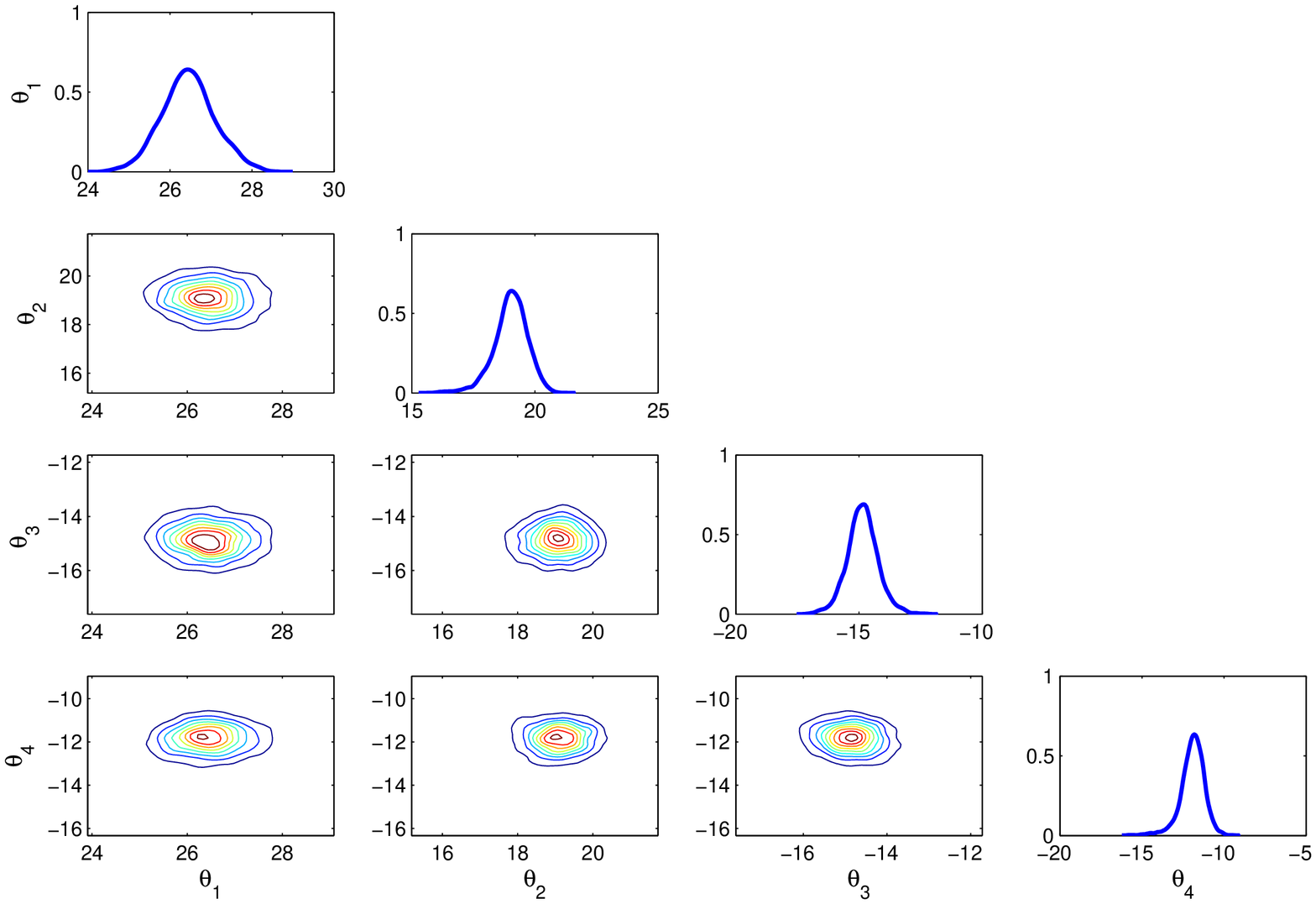}
\caption{One dimensional and two dimensional posterior marginals of the largest DCT coefficients for the simulated model.}
\label{contourex1}
\end{figure}

Next, we consider the permeability field generated by GEO-R as a reference spatial field, which is treated as the unknown spatial field in our model. The observed coarse-scale log permeability field, $\bm{y_c}$, is calculated using the upscaling procedure in a $5 \times 5$ coarse grid. So the scale difference of the coarse-scale permeability field with respect to the fine-scale permeability field is $5$ unit in each direction, i.e, each grid on coarse-scale data is of $0.2 \times 0.2$ square unit. The fine-scale log permeability field is assumed to be observed only at $6$ well locations along the boundaries, denoted by $\bm{y_o}$. We use $50$ observed real output data corresponding to the reference permeability field and $5000$ simulation data corresponding to the $100$ simulation runs in our model. We simulate $200000$ samples from the posterior by the hybrid reversible jump MCMC, Gibbs, and Metropolis-Hastings method described before. After the $10000$ burn-in period we retain every $10$th sample. Figure \ref{emuperm2} shows the reference log permeability field and the mean of the posterior log permeability field. We can see that the posterior mean is very close to the reference log permeability field. Figure \ref{simupdf1} shows the posterior density of the highest DCT coefficient, $\theta_1$ and the posterior density of $\sigma_{z}^2$. The posterior density of $\theta_1$ has a peak near $26$ which is close to the true highest DCT coefficient($26.43$) of the transformed reference field. The marginal one-dimensional and two-dimensional posterior distribution of the top four DCT coefficients are shown in Figure \ref{contourex1}.
We also compare the computational efficiency in terms of CPU time for the emulator-based MCMC method to the regular simulator-based MCMC method. The results from Table \ref{tabcomp} show that the emulator-based method is at least $20$ times faster than the direct simulator-based MCMC method.

\begin{table}[h!]
\begin{center}
\begin{tabular}{|c||p{4cm}|p{4cm}| p{4cm}|}
\hline 
MCMC method &Time per likelihood calculation & Time per MCMC iteration & Total time for inversion (200000 samples)\\
\hline
Simulator based & 5.000 & 5.112 &  1022400 \\
Emulator based & 0.112 & 0.212 &  42900  \\
\hline
\end{tabular}
\end{center}
\caption{Computational times, in seconds, of the emulator based and simulator based MCMC methods.} \label{tabcomp}
\end{table}

\subsection{Numerical results for a real field example}

\begin{figure}[h!]
\centering
\includegraphics[width=6.5in,height=4.5in]{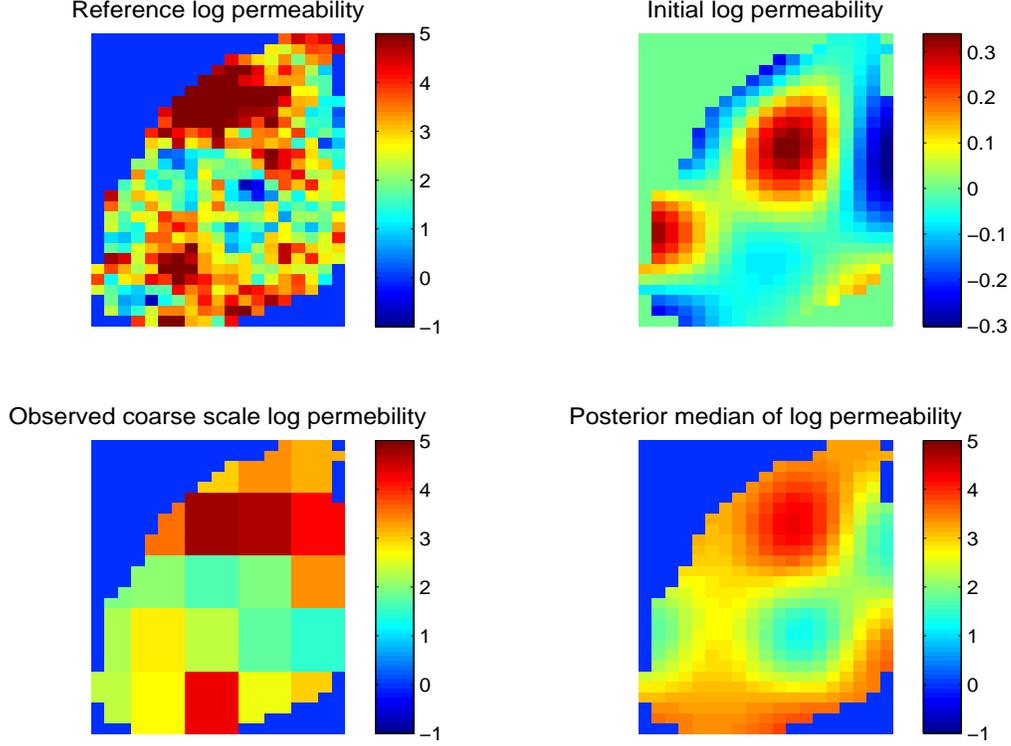}
\caption{Log permeability fields for the PUNQ-S3 model. Top left: Reference field, Top right: Initial field of the Markov chain, Bottom Left: Observed coarse-scale permeability, Bottom right: Posterior median.}
\label{punqpermbmars1}
\end{figure}

\begin{figure}[htbp]
\centering
\includegraphics[width=4in,height=3.3in]{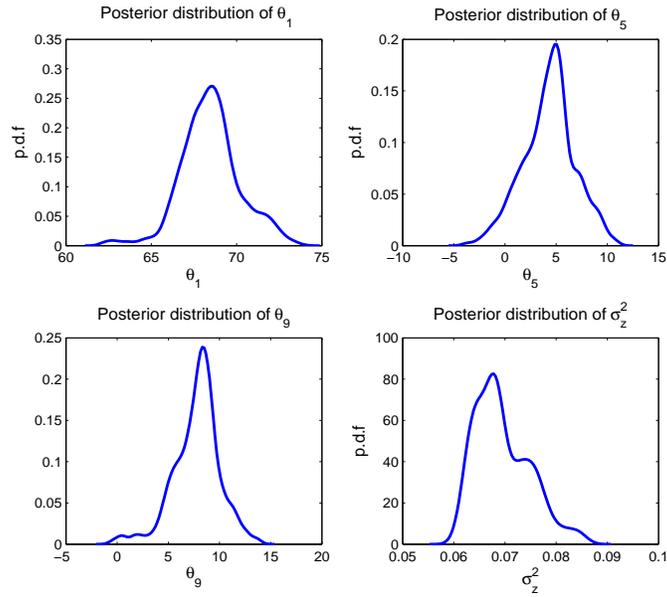}
\caption{Posterior distributions for the PUNQ-S3 model. Top Left: Posterior density of $\theta_1$, Top Right: Posterior density of $\theta_5$, Bottom Left: Posterior density of $\theta_9$, Bottom Right: Posterior density of $\sigma_{z}^{2}$.}
\label{punqpdfbmars1}
\end{figure}

\begin{figure}[htbp]
\centering
\includegraphics[width=5.5in,height=3.8in]{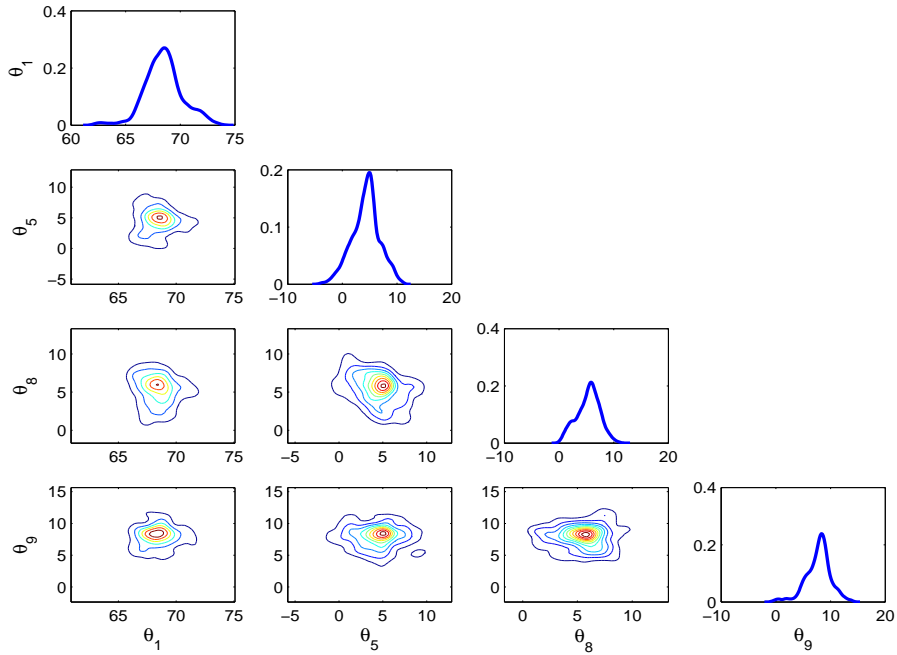}
\caption{One dimensional and two dimensional posterior marginals of the largest DCT coefficients for the PUNQ-S3 model.}
\label{contourpunq1}
\end{figure}

\begin{figure}[htbp]
\centering
\includegraphics[width=3.5in,height=2.5in]{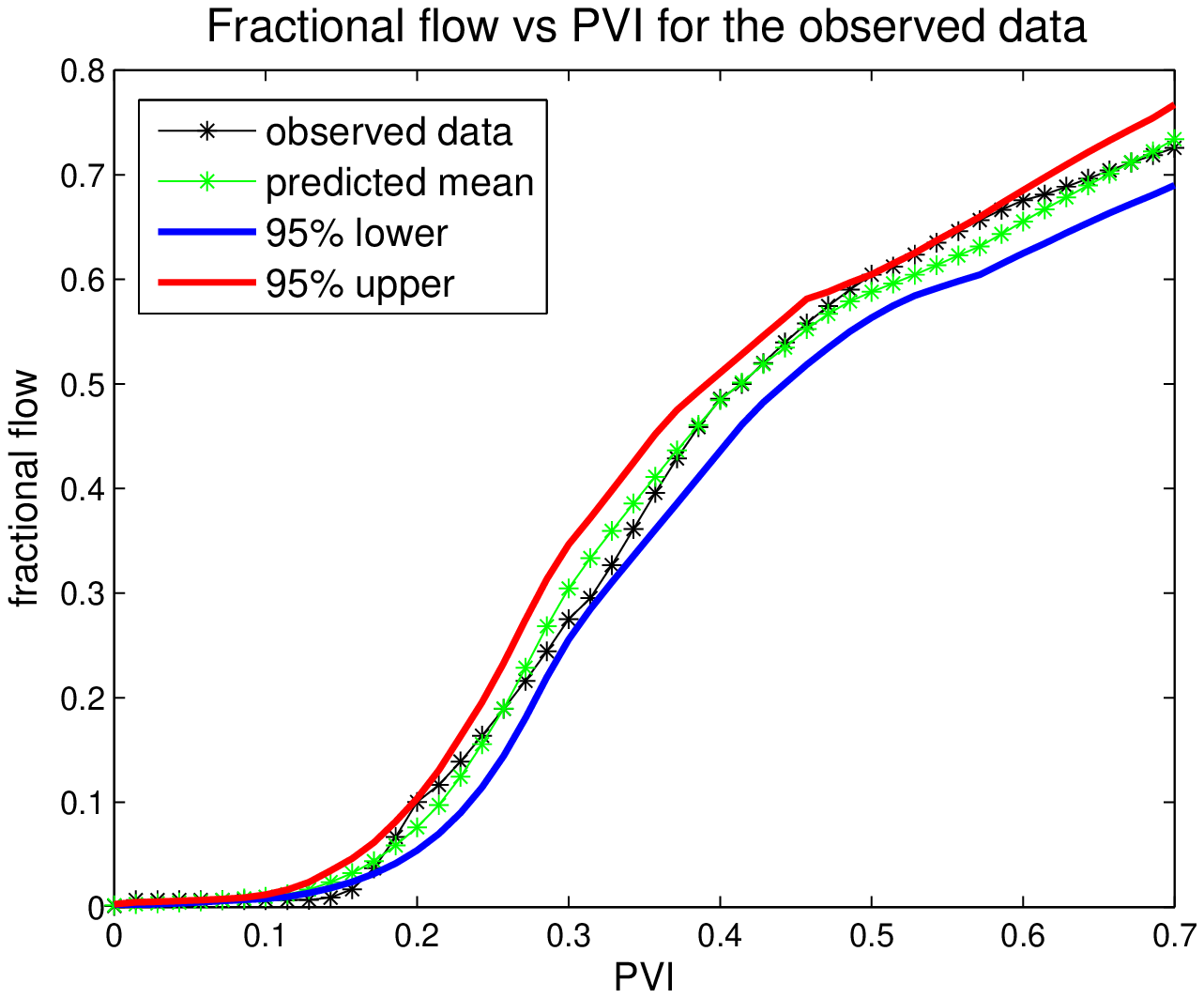}
\caption{Fitted mean and $95\%$ credible interval of the observed output for the PUNQ-S3 model using the emulator}
\label{fittedpunq1}
\end{figure}

Here we apply our model on a real field example, viz., PUNQ-S3 data-set. The PUNQ-S3 model has been taken from a reservoir engineering study on
a real field example provided by Elf Exploration Production. It was qualified as a small-size industrial reservoir engineering model. The PUNQ-S3 data set was an experimental study where the true permeability was actually known on the $19 \times 28 \times 5$ grid but the researchers were asked not to use the permeability data for their modeling purpose. They were asked to use the production history only to infer about the true permeability field and then compare how their model resembles the true permeability field. For our example, we consider only the second top layer of the
five layers in the model and follow the same guidelines. We have used the $50$ production
history i.e., the water-cut data, the permeability data on a $5 \times 5$ coarse grid, and the true fine-scale permeability data only on the $6$ well locations to infer about the fine-scale permeability field. The permeability measurements are expressed in the unit of mD where $1mD = 10^{-3}$ Darcy = $10^{-12}m^2$. The spatial locations of the fine-scale permeability field were given to the researchers in a transformed Cartesian coordinate system with each grid of $180\times180$ square unit starting from the origin, i.e., the coordinate of the top-left grid block is $(0,0)$ and that of the bottom-right grid block is $(3420, 5040)$. Each grid on coarse-scale data is of $684 \times 1008$ square unit. We use log transformation of the permeability
data and logit transformation of the fractional flow data in our model. To build the BMARS emulator we generate $100$ samples of $16$ DCT coefficients using Latin hypercube sampling from multivariate Normal distribution. Each of these sets of $16$ DCT coefficients corresponds to a log permeability field obtained by the respective inverse DCT transformation. The other type of input considered is $50$ injected pore volumes (rescaled to $0-1$) for each of these spatial fields. For each of these $5000$ simulated input observations the output or water-cut data is simulated using computer codes for the forward simulator. We use $50$ observed real water-cut data and $5000$ simulated water-cut data in our model. We draw $200000$ samples from the posterior distribution, after the $10000$ burn-in period we retain every $10$th sample. From Figure \ref{punqpermbmars1} we can see that the posterior median of the sampled permeability field is close to the reference permeability field. The marginal posterior distribution of some of the model parameters is shown in Figure \ref{punqpdfbmars1}. Even though we have used almost flat priors for the model parameters, from the posterior marginals we can see that the observed data can reduce the uncertainties of the model parameters. The marginal one-dimensional and two-dimensional posterior distribution of the top four DCT coefficients are shown in Figure \ref{contourpunq1}. From the marginal distribution, we can see that the marginal posterior for all the DCT coefficients have a peak near the true value of DCT coefficients obtained by the DCT transformation of the reference log permeability field. Hence we can conclude that our Bayesian model can quantify the uncertainties in the unknown permeability field very well. The mean of the fitted output data corresponding to the reference permeability field together with its $95\%$ credible interval is shown in Figure \ref{fittedpunq1}. From the predictive mean and the credible interval, we can conclude that the BMARS emulator predicts the output very well.
\section{Conclusion}

The paper pursues a Bayesian approach to inverse problems in which the unknown quantity is a spatial field. The posterior distribution provides a quantitative assessment of the uncertainty in the inverse solution. The computational challenges associated with the repeated evaluation of the forward simulator are addressed. We use emulators based on the Bayesian approach to multivariate additive regression splines to avoid the computational challenges of the direct simulation-based approach. The unknown spatial field is parameterized by DCT transformation and the transformed DCT coefficients are used as regressors in the BMARS model. Numerical results show that the BMARS emulator-based MCMC method has substantial efficiency gain over the simulator-based MCMC method in terms of CPU time. Our method is very flexible and can be applied to any physical process whose input is a spatial field. The method can be adapted to other inverse problems very easily as the mathematical model for the physical process was never used in the model. Moreover, the developed BMARS emulators can be easily used for prediction purposes which is very important in many fields such as production forecasting in oil reservoirs.
%


\bibliography{Emulator_Reference}


\end{document}